# From P&ID Drawings to Process Graphs: A Multimodal Language Model Approach

**Baikai Zhu[a], Samuel Duong[a], Javal Vyas[a,*], Mehmet Mercangöz[a]**

[a] Imperial College London, Department of Chemical Engineering, London, Greater London, United Kingdom
* Corresponding Author: j.vyas24@imperial.ac.uk

## ABSTRACT

Piping and instrumentation diagrams (P&IDs) encode the functional structure of process plants and are a critical yet underutilised source of engineering knowledge for digital twins and intelligent decision support. However, digitising legacy P&IDs remains challenging due to heterogeneous drawing standards and the reliance of existing methods on brittle symbol recognition and rule-based connectivity reconstruction. This work reframes P&ID digitization as the extraction of equipment tags and inference of process topology, rather than graphical reproduction. We propose a two-stage workflow based on multimodal large language models, in which visual extraction and topology reconstruction are treated as distinct reasoning stages guided by chemical engineering process knowledge. The approach is evaluated on two ANSI-standard P&ID case studies of increasing complexity. Results show that decomposing visual extraction and topology reasoning yields more accurate and structurally consistent process representations than end-to-end digitization, highlighting the potential of language-model-based, knowledge-guided workflows for scalable and semantically reliable P&ID digitization.

**Keywords**: P&ID digitisation, Multimodal large language models, Graph reconstruction

## INTRODUCTION

Piping and instrumentation diagrams (P&IDs) constitute the structural backbone of a process plant, capturing how equipment, instrumentation, and control elements are interconnected to realise the intended process operation. Beyond providing spatial or locational information, P&IDs encode the functional logic of the plant, describing how units interact, how material and signals propagate, and how control strategies are implemented. As such, they represent an authoritative source of process understanding that spans design intent, operational constraints, and safety considerations.

Despite their central role, the information embedded in P&IDs remains largely underexploited in industrial practice. In many plants, P&IDs continue to be treated primarily as static reference documents rather than as structured, machine-interpretable sources of knowledge. This limits their integration into advanced engineering workflows, where explicit representations of process topology could support automated consistency checking, decision support, and interaction with digital twins. As the process industries move toward smart factories and data-driven operation, the ability to interrogate and reason over P&IDs becomes increasingly important.

A major barrier to such “smart” use of P&IDs is the legacy nature of existing documentation. A substantial fraction of industrial P&IDs still exists as paper drawings or unstructured raster images, often produced over decades using heterogeneous drafting standards and conventions. Converting these legacy artefacts into structured, machine-readable representations remains labour-intensive and typically requires extensive manual interpretation by domain experts. This manual effort does not scale and represents a significant obstacle to unlocking the full value of P&IDs within modern digital infrastructures, including retrieval-augmented generation systems and interactive engineering tools.

From a technical standpoint, P&ID digitization is challenging due to variability in symbol rendering, annotation practices, and drawing quality across companies and engineering contractors. While human engineers resolve such variability using contextual and domain knowledge, automated systems must operate on inconsistent and ambiguous graphical representations. Conventional digitization pipelines therefore rely on deep learning models for symbol and text recognition combined with rule-based methods to reconstruct process connectivity [1–3]. Although effective in constrained settings, these approaches exhibit limited generalisation across drawing standards [1,5], require extensive manual engineering of reconstruction rules [4], and propagate errors from early recognition stages into the reconstructed topology. Consequently, many existing methods emphasise graphical detection while falling short of producing semantically reliable representations of process structure.

Recent advances in vision-language models (VLMs) and multimodal large language models (MLLMs) provide an opportunity to reconsider this paradigm. By jointly grounding visual and semantic information, such models can reduce dependence on specific symbol appearances and bespoke training for individual drawing standards. When combined with language-based reasoning, they further enable inference over extracted engineering artefacts in a manner that more closely resembles expert interpretation of P&IDs.

In this work, we investigate the use of multimodal language models and multi-agent workflows to transform legacy P&IDs into structured, semantically meaningful representations of process topology. By decoupling visual extraction from topology reasoning, the proposed approach aims to reduce manual digitization effort and enable P&IDs to function as active, machine-interpretable sources of process knowledge, supporting intelligent interaction and decision-making in future smart manufacturing environments.

# LITERATURE REVIEW

Existing research on P&ID digitization has primarily focused on automating the extraction of symbols, text, and connectivity from legacy diagrams using computer vision and machine learning techniques. The following review summarises the main methodological directions, highlighting their strengths and limitations with respect to robustness, scalability, and semantic correctness.

## Classical image processing and rule-based pipelines

Early approaches to P&ID digitization predominantly relied on classical image processing, optical character recognition (OCR), and engineering heuristics. These systems typically detect graphical primitives such as symbols, lines, and text using template matching and rule-based algorithms, followed by proximity-based association to identify equipment tags and infer connectivity. Kang et al. [6] demonstrated the feasibility of converting raster P&IDs into structured databases using such techniques, while maintaining compatibility with downstream engineering tools. However, these methods are highly sensitive to drawing quality, symbol variability, and vendor-specific drafting conventions, which limits their robustness when applied to heterogeneous industrial datasets [7].

## Deep learning for visual recognition

To address the limitations of purely rule-based methods, subsequent research has integrated deep learning for visual recognition tasks. Convolutional neural networks and modern text detection architectures have been applied to identify equipment symbols, instrumentation, and tag text in P&IDs [3]. Industrially oriented pipelines combine deep learning–based detection of symbols, text, and lines with heuristic association rules to reconstruct connectivity and export structured representations aligned with engineering information models [8]. While these approaches significantly improve robustness in component recognition, topology reconstruction remains challenging, as rule-based connectivity inference is sensitive to broken lines, ambiguous junctions, and overlapping annotations commonly found in real plant drawings.

## Graph-based formulations of topology reconstruction

Recognising that the primary engineering value of a P&ID lies in its process topology, several studies explicitly model digitization as a graph reconstruction problem. In this formulation, equipment and instruments form graph nodes, while piping and signal connections define edges. Mani et al. (2020) [9] combine deep learning–based detection with graph search techniques to traverse line structures and infer connectivity. Similarly, Digitize-PID introduces a structured pipeline that incorporates domain-based validation to correct implausible connections and improve graph consistency [1]. More recently, relation-based and transformer-style models have been proposed to directly infer connectivity from diagram images, framing P&ID understanding as an image-to-graph task [9]. Although these methods show improved structural reasoning, they typically require specialised training data with explicit edge annotations and may struggle to generalise across diverse drafting styles.

## Data scarcity and generalisation challenges

A practical challenge in P&ID digitization is the scarcity of annotated datasets, as industrial drawings are often proprietary. To mitigate this, recent work explores semi-supervised and weakly supervised learning strategies. Gupta et al. (2024) [11] propose a semi-supervised symbol detection approach that reduces annotation effort while improving generalisation across symbol variants. Complementary methods exploit auxiliary engineering artefacts such as legend sheets to infer symbol semantics with minimal manual labelling [12]. While these approaches improve robustness in symbol recognition, they do not directly address the inference of process connectivity under uncertainty.

### Engineering knowledge and validation

Across the literature, it is widely acknowledged that purely visual or data-driven methods are insufficient to guarantee engineering correctness. Equipment type, tag nomenclature, and process conventions impose strong constraints on allowable connectivity—for example, pumps require suction and discharge connections, vessels connect via nozzles, and instruments interface through signal rather than process lines. Several studies therefore incorporate post-processing validation stages based on engineering rules to filter implausible structures and enforce consistency with expected process behaviour [8,9]. However, such rules are typically applied locally and procedurally, limiting their ability to resolve global ambiguities in complex diagrams.

### Language-based reasoning for P&ID understanding

Recent advances in large language models (LLMs) and vision-language models (VLMs) have motivated exploratory studies on their application to engineering diagrams. Initial investigations indicate that direct application of LLMs to P&ID digitization without grounding or constraint enforcement leads to inconsistent or non-physical interpretations [13]. Nevertheless, these models demonstrate strong capabilities in reasoning over structured engineering information when provided with explicit inputs and constraints, suggesting their potential role in topology inference when combined with reliable visual extraction and engineering knowledge.

## METHODOLOGY

### Problem formulation

We adopt a process-centric formulation of the P&ID digitization problem. Rather than aiming to reproduce graphical details of the diagram, the primary objective is to extract (i) identifiable process units and instrumentation, represented by their tags, and (ii) the interconnections between these units that define the process topology. Under this formulation, a digitised P&ID is represented as a graph, where nodes correspond to equipment or instrumentation tags and edges represent material, signal, or control connections between them.

Conventional digitization pipelines typically address this task through a combination of deep learning–based visual recognition and rule-based connectivity reconstruction. In contrast, we investigate whether modern multimodal large language models (MLLMs), combined with language-based reasoning, can directly infer structured process representations from P&ID images, and whether decomposing the task into extraction and reasoning stages improves robustness and accuracy.

### Overview of evaluation pathways

We assess language-model-based P&ID digitization using two complementary workflows, illustrated schematically in Figure 1. The first workflow performs extraction and topology reconstruction in a single step using an MLLM. The second workflow decomposes the task into two stages—node extraction and topology reconstruction—handled by separate language model instances, with the reconstruction stage augmented by explicit chemical engineering process knowledge. In both cases, the final output is a structured graph representation of the P&ID, which is evaluated against a manually constructed ground truth.

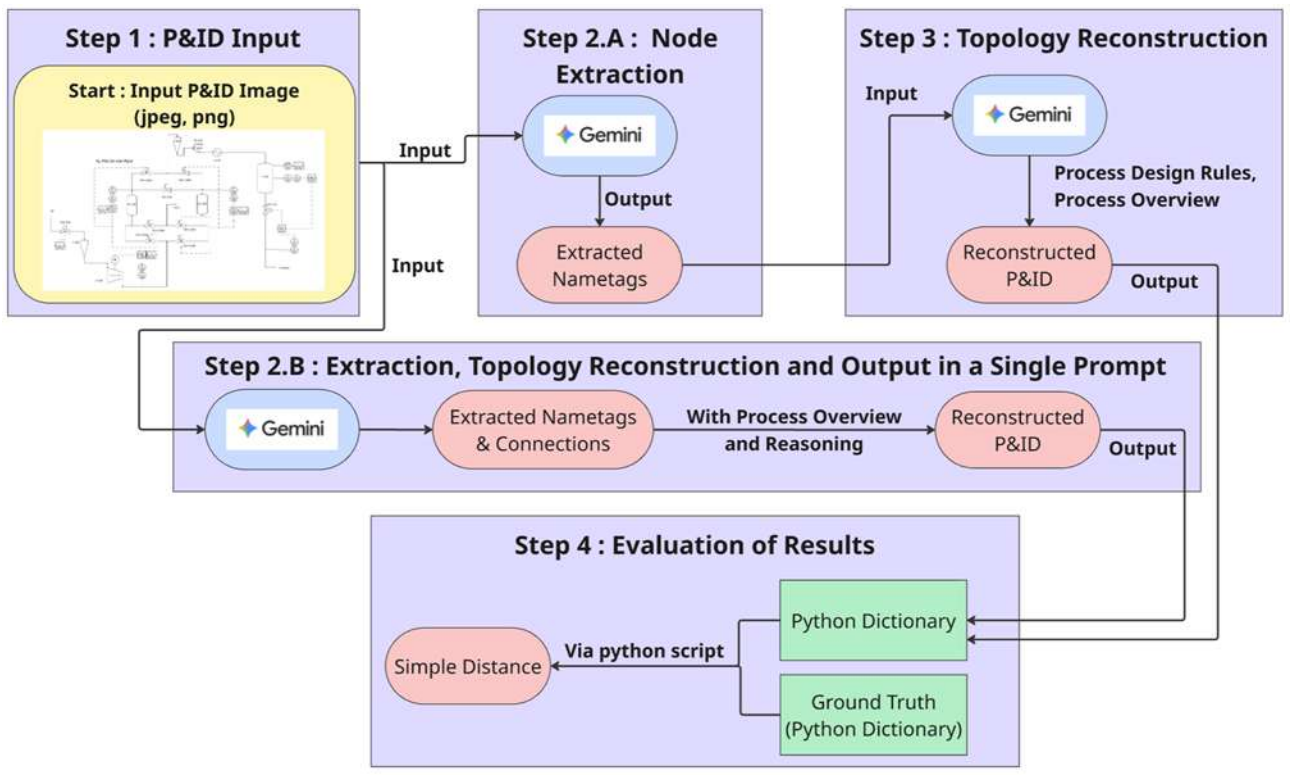

**Figure [1]** Overview of the proposed P&ID digitisation workflows

### Pathway I: End-to-end digitization with an MLLM

In the first workflow (Figure 1, Step 2.B), the P&ID image is divided into eight spatial slices to mitigate resolution and context-length limitations of the multimodal model. These image slices, together with a short high-level textual description of the diagram, are provided as input to an MLLM. The model is prompted to digitise the P&ID directly into a structured representation expressed as a Python dictionary, encoding process units as nodes and their interconnections as edges.

To assess the quality of the generated representation, a ground-truth graph of the same P&ID is manually constructed using the same dictionary format. The predicted and ground-truth graphs are compared using three metrics. Node identification accuracy measures the proportion of correctly identified tags, while edge recognition accuracy measures the proportion of correctly inferred connections. In addition, overall structural similarity is quantified using a simple distance metric. The simple distance is defined as the total number of node- and edge-level discrepancies between the predicted and ground-truth graphs, including missing nodes, extra nodes, missing edges, extra edges, and edge-type mismatches.

This workflow evaluates the feasibility of treating P&ID digitization as a single-step multimodal reasoning task.

## Pathway II: Decomposed extraction and topology reconstruction

The second workflow (Figure 1, Steps 2.A and 3) explicitly separates visual extraction from topology inference, reflecting the hypothesis that reasoning over process structure benefits from abstraction away from raw visual input.

### Node extraction

In the extraction stage (Figure 1, Step 2.A), an MLLM is provided only with the P&ID image slices and tasked with identifying the set of process units and instrumentation present in the diagram, essentially functioning as an advanced Optical Character Recognition (OCR) tool. The model outputs a list of equipment and instrument tags without attempting to infer connectivity, thereby isolating the visual recognition task.

### Topology reconstruction with engineering knowledge

In the reconstruction stage (Figure 1, Step 3), the output of the extraction step is passed to a second language model. This model has no access to the original P&ID images and relies exclusively on symbolic inputs. In addition to the extracted node list, the model is provided with (i) a high-level process overview and (ii) a document describing general chemical engineering process design rules. It is important to note that the inclusion of such chemical engineering context does not involve fine-tuning or training the underlying model. Instead, this study utilises an off-the-shelf MLLM (Gemini-3.0-Pro), where domain-specific knowledge is introduced entirely via prompt engineering, supplied to the language model in the form of an accompanying text document.

The process design rules are organised into three categories. The first category defines common tag nomenclature and abbreviations for equipment, valves, and instrumentation, capturing typical conventions used to encode process streams, destinations, and unit sequencing. The second category describes common physical connectivity patterns between process units, reflecting standard design and maintenance practices. The third category addresses control and instrumentation structure, guiding the model in inferring control loops and signal hierarchies, including the directionality of signals between sensors, controllers, and actuators.

Using these inputs, the reconstruction model infers the connectivity between nodes and outputs a structured graph representation in the same format as Workflow I.

### Evaluation Protocol

For both workflows, the reconstructed P&ID graph is evaluated against the same ground-truth representation using identical metrics, as shown in Figure 1 (Step 4). This enables a direct comparison between end-to-end multimodal digitization and the decomposed extraction–reasoning approach, isolating the effect of task decomposition and explicit process knowledge on topology reconstruction accuracy.

# CASE STUDY

To evaluate the proposed P&ID digitization workflows, two representative process diagrams were selected as case studies. Both P&IDs were designed in accordance with the ANSI standard and use consistent naming conventions for equipment and instrumentation to minimise variability arising from tag nomenclature rather than diagram structure.

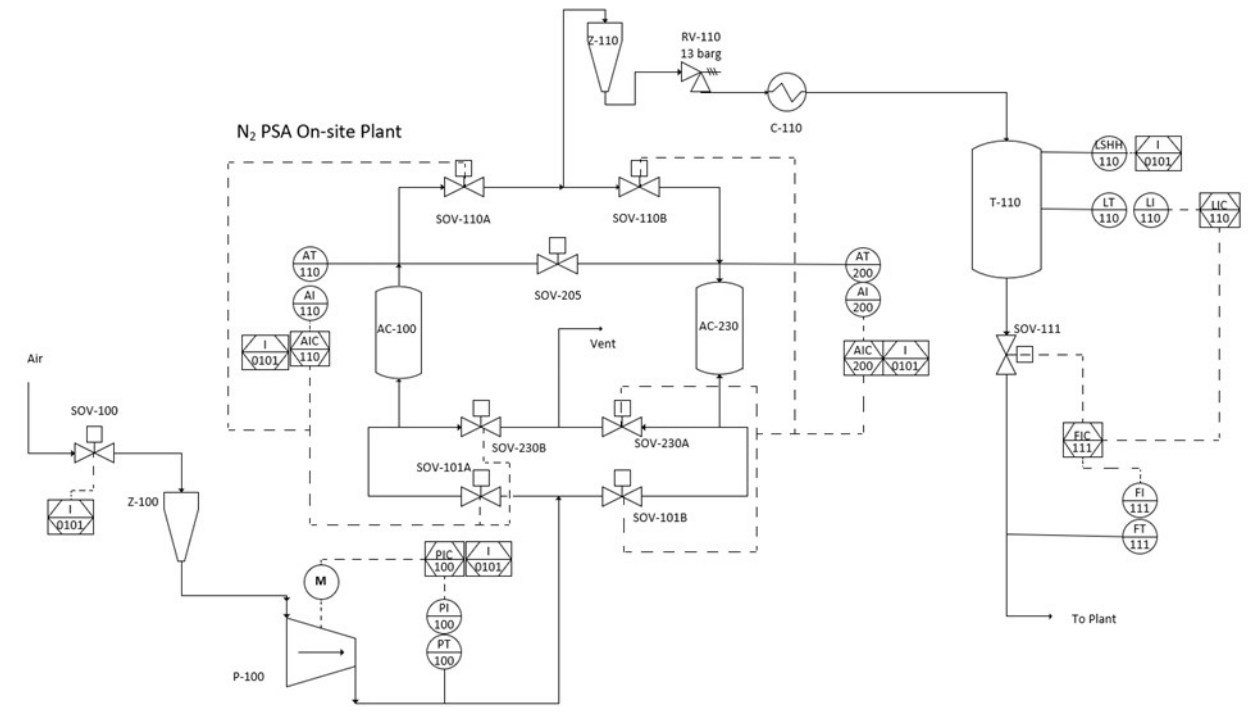

**Figure [2]** Nitrogen Pressure Swing Adsorption P&ID

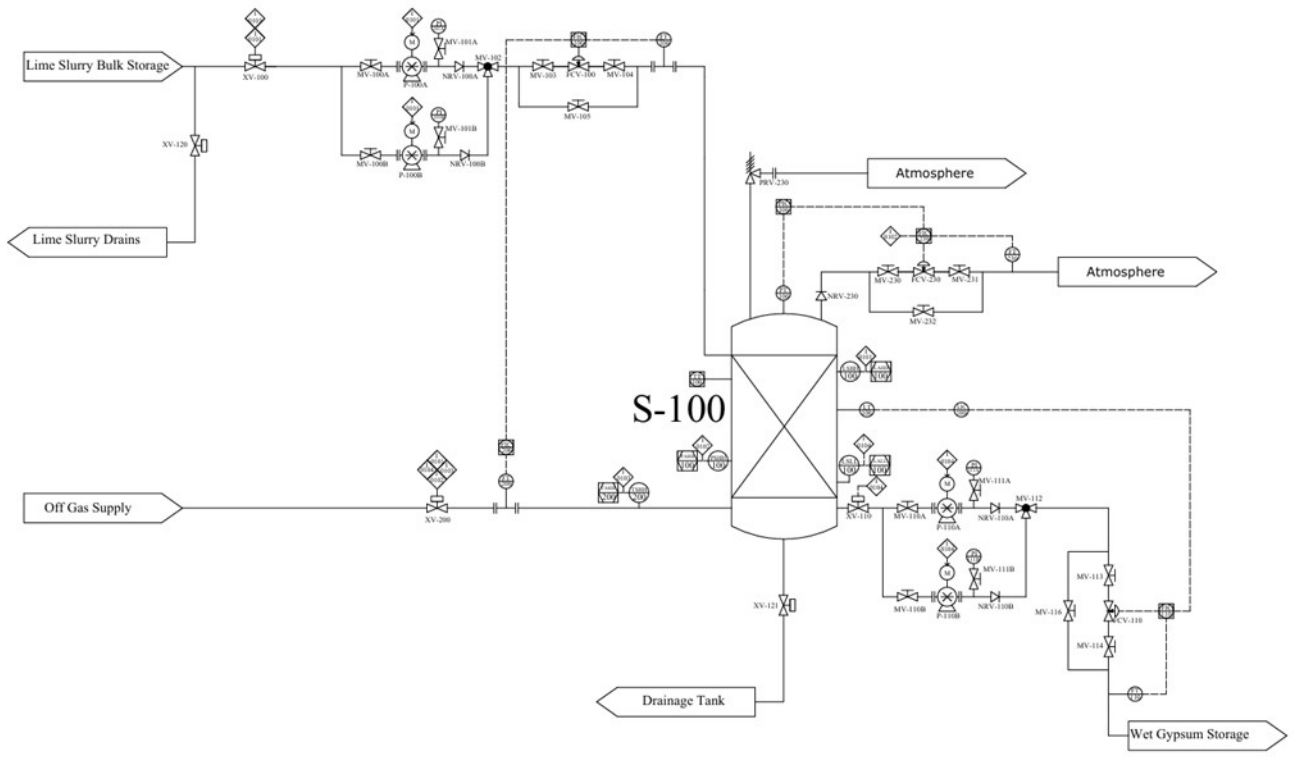

**Figure [3]** Wet Flue Gas Desulfurization P&ID

The first case study, shown in **Figure 2**, corresponds to a nitrogen pressure-swing adsorption (PSA) process. The second case study, shown in **Figure 3**, represents a wet flue gas desulfurization (FGD) process. These two systems were chosen to reflect different levels of process and topological complexity while remaining representative of industrial process flows commonly encountered in practice.

## P&ID complexity

We quantified the complexity of each P&ID by comparing the total number of nodes and edges, where a higher number of nodes and edges represents a more complex case study, as shown in Table [1].

**Table [1]** P&ID Complexity Metrics

| P&ID | Complexity |
|---|---|
| Nitrogen Pressure Swing Adsorption | Total nodes: 38<br>Total edges: 53<br>Total combined: 91 |
| Wet Flue Gas Desulfurization | Total nodes: 73<br>Total edges: 101<br>Total combined: 174 |

## Experimental setup

At the start of each case study, the P&ID image was divided into eight equal-resolution slices to mitigate resolution and context-length limitations of the multimodal language models. In addition, a short, high-level textual description of the process was prepared for each P&ID to provide contextual information where applicable.

For the end-to-end workflow, the P&ID slices and process overview were provided to Gemini-3.0-pro in a single prompt. The model was instructed to output a Python dictionary representing the P&ID, with equipment and instrumentation tags as nodes and their interconnections as edges.

For the decomposed workflow, digitization was split into two stages using a multi-agent setup implemented in Dify, an agentic workflow framework. In the first stage, the language model was provided only with the P&ID image slices and tasked with extracting equipment and instrumentation tags without inferring connectivity. The model is prompted to solely extract equipment tags and numbering while ignoring the graphical symbols entirely. In the second stage, the extracted node list was passed to a separate instance of the language model, together with the process overview and a document describing general chemical engineering process design rules. This second instance inferred the interconnections between nodes using language-based reasoning on the linguistic meaning of the extracted tags, guided by chemical engineering domain knowledge regarding standard configurations of process design. Since no graphical symbols are used as an input to this reconstruction stage, the traditional challenge of spatially mapping text annotations to corresponding visual symbols is bypassed. Importantly, the reconstruction model had no access to the original P&ID images and shared no memory with the extraction model beyond the list of identified nodes.

Although Gemini-3.0-pro was used for both extraction and reconstruction in this case study, the workflows are model-agnostic. Any vision-language model can be substituted in the extraction stage, and any reasoning-capable language model can be used in the reconstruction stage.

## Evaluation procedure

For each P&ID case study, a ground-truth representation was manually constructed in the same Python dictionary format as the model outputs, with equipment and instrumentation tags represented as nodes and their interconnections represented as edges. Performance was evaluated by comparing the predicted graph against the ground truth using accuracy-based metrics and a structural discrepancy metric.

To account for stochastic variability in language model outputs, each workflow was executed ten times using identical prompts. All reported results correspond to the mean and standard deviation across these runs.

## Comparative Analysis

The experimental design enables several comparative analyses. First, the performance of end-to-end digitisation with an MLLM (E2E-MLLM) is compared against decomposed extraction and topology reconstruction (DETR) to assess the effect of task decomposition. Second, the end-to-end workflow is evaluated with and without a high-level process overview to examine the impact of contextual information. Finally, within the decomposed workflow, topology reconstruction with and without explicit chemical engineering design rules is compared to quantify the contribution of domain-specific knowledge to connectivity inference.

Together, these comparisons isolate the roles of contextual information, task decomposition, and engineering knowledge in language-model-based P&ID digitisation.

# RESULTS

The results reported in Tables 2.A and 2.B are evaluated using three accuracy-based metrics and one structural discrepancy metric. Total accuracy measures the proportion of correctly reconstructed nodes and edges relative to the ground truth, while node accuracy and edge accuracy quantify the correctness of equipment identification and connectivity inference, respectively.

In addition to these local accuracy measures, global structural similarity between the predicted and ground-truth graphs is quantified using a simple distance metric. This metric counts the total number of discrete modifications required to reconcile the predicted graph with the ground truth, including missing nodes, extra nodes, missing edges, extra edges, and edge-type mismatches. A lower simple distance therefore indicates that fewer structural changes are required for the reconstructed

P&ID to match the reference topology, providing a direct and interpretable measure of topological correctness. All reported values are expressed as mean ± standard deviation over ten independent runs.

Formally, the simple distance is defined as:

$$D_s = |V_{miss}| + |V_{extra}| + |E_{miss}| + |E_{extra}| + |E_{type}| \quad (1)$$

where $V_{\text{miss}}$ and $V_{\text{extra}}$ denote missing and extra nodes, $E_{\text{miss}}$ and $E_{\text{extra}}$ denote missing and extra typed edges, and $E_{\text{type}}$ captures edge-type mismatches between the predicted and ground-truth graphs.

The accuracy metrics are computed as follows:

$$\textit{Total Accuracy} = \frac{\sum(\textit{Correct Nodes} + \textit{Correct Edges})}{\sum(\textit{Total Nodes} + \textit{Total Edges})} \quad (2)$$

$$\textit{Node Accuracy} = \frac{\sum \textit{Correct Nodes}}{\sum \textit{Total Nodes}} \quad (3)$$

$$\textit{Edge Accuracy} = \frac{\sum \textit{Correct Edges}}{\sum \textit{Total Edges}} \quad (4)$$

## Nitrogen PSA plant

Results for the nitrogen pressure-swing adsorption (PSA) P&ID are summarised in Table 2.A. Using end-to-end digitization with a multimodal large language model (E2E-MLLM) as the baseline configuration, the model achieves a total accuracy of 68.56%, with relatively high node accuracy (86.32%) but substantially lower edge accuracy (55.58%). This disparity indicates that while equipment identification is largely successful, connectivity inference remains challenging when extraction and reasoning are performed jointly.

Augmenting the end-to-end prompt with a process overview yields a modest improvement across all metrics, increasing total accuracy to 70.67% and reducing the simple distance from 45.6 to 41.0. This suggests that high-level contextual information helps resolve some connectivity ambiguities; however, structural discrepancies remain pronounced, as reflected by the still relatively high simple distance.

A more substantial improvement is observed when switching to the decomposed extraction and topology reconstruction (DETR) workflow augmented with both a process overview and explicit process design rules. In this configuration, DETR achieves a markedly higher total accuracy of 88.78%, with node accuracy reaching 100% and edge accuracy improving to 80.58%. This improvement is accompanied by a significant reduction in simple distance to 24.0, indicating that substantially fewer node and edge modifications are required to align the reconstructed graph with the ground truth.

## Slime and Off gas Plant

Results for the slime and off-gas plant P&ID, which represents a more complex process topology, are shown in Table 2.B. In this case, the limitations of end-to-end digitisation are more pronounced. The baseline E2E-MLLM configuration achieves a total accuracy of 67.31% and exhibits a high simple distance (68.20), indicating substantial structural deviation from the reference topology.

Providing a process overview in the end-to-end workflow improves total accuracy to 71.81% and reduces the simple distance to 59.70, highlighting the increasing importance of contextual information as diagram complexity grows. Nevertheless, a significant number of structural discrepancies persist, particularly in connectivity inference.

Transitioning to the decomposed extraction and topology reconstruction (DETR) workflow further clarifies the role of domain knowledge. While decomposed extraction and reconstruction without explicit process knowledge does not consistently outperform end-to-end digitisation, the highest performance is achieved when DETR is augmented with both a process overview and chemical engineering design rules. In this configuration, the model attains a total accuracy of 85.21%, near-perfect node accuracy (99.73%), and a substantially reduced simple distance of 48.30, indicating a closer match to the ground-truth process topology.

**Table [2.A] Nitrogen PSA Plant Results**

| | E2E | E2E + Knowledge | DETR | DETR + Knowledge |
|---|---|---|---|---|
| Total Acc.(%) | 68.56 ± 0.07 | 70.67 ±0.06 | 71.22 ± 0.03 | 88.78 ± 0.02 |
| Node Acc. (%) | 86.32 ± 0.07 | 87.37 ± 0.05 | 100.00± 0.00 | 100.00 ± 0.00 |
| Edge Acc. (%) | 55.58 ± 0.08 | 59.98 ± 0.09 | 50.19 ± 0.05 | 80.58 ± 0.04 |
| Simple Distance | 45.60 ± 6.52 | 41.00 ± 6.70 | 51.80 ± 3.99 | 24.00 ± 4.67 |

**Table [2.B] Slime & Offgas Plant Results**

| | E2E | E2E + Knowledge | DETR | DETR + Knowledge |
|---|---|---|---|---|
| Total Acc. (%) | 67.31 ± 0.07 | 71.81 ± 0.06 | 64.74 ± 0.02 | 85.21 ± 0.03 |
| Node Acc. (%) | 78.38 ± 0.08 | 78.20 ± 0.09 | 100.00± 0.00 | 99.73 ± 0.01 |
| Edge Acc. (%) | 58.87± 0.07 | 66.63 ± 0.07 | 37.84± 0.04 | 74.12 ± 0.05 |
| Simple Distance | 68.20 ± 12.64 | 59.70± 10.90 | 116.60 ± 8.22 | 48.30 ± 10.23 |

Across both case studies, several consistent trends emerge. First, DETR augmented with process knowledge consistently outperforms end-to-end digitisation across all metrics, with the largest gains observed in edge accuracy and simple distance. Second, providing high-level process context improves performance for both workflows but does not fully resolve structural inconsistencies when extraction and reasoning are coupled. Third, incorporating explicit chemical engineering design rules during topology reconstruction yields the most structurally faithful representations, as evidenced by the lowest simple distances.

Taken together, these results show that accurate P&ID digitisation requires not only reliable visual extraction but also explicit reasoning over process structure, and that separating these concerns—when guided by domain knowledge—leads to reconstructions that are both more accurate and more closely aligned with engineering intent.

## CONCLUSION

This work investigated the use of multimodal large language models and decomposed, agent-based workflows for the semantic digitisation of piping and instrumentation diagrams. By reframing P&ID digitisation as a problem of extracting equipment tags and inferring process topology, we demonstrated that separating visual extraction from topology reconstruction leads to more accurate and structurally consistent representations than end-to-end digitisation with a single prompt. Across two industrially representative case studies, the proposed decomposed extraction and topology reconstruction workflow consistently outperformed end-to-end approaches, particularly in edge accuracy and structural consistency.

The results show that incorporating high-level process context and explicit chemical engineering design rules during topology reconstruction substantially improves performance, especially for more complex diagrams. This highlights the importance of embedding domain knowledge directly into the digitisation workflow, allowing chemical engineers to actively contribute to the definition of process rules rather than relying on extensive manual refinement of rule-based or data-driven computer vision systems. The resulting structured representations align naturally with graph-based standards such as DEXPI, supporting downstream integration with digital twins and intelligent engineering tools.

A practical limitation of the proposed approach is the use of proprietary language models, which raises confidentiality concerns when handling sensitive industrial P&IDs. However, the workflow itself is model-agnostic and can be readily adapted to open-source multimodal language models, enabling deployment in environments with strict data governance requirements. A systematic comparison between proprietary and open-source models was beyond the scope of this study and remains an important direction for future work.

Future research will focus on closing the loop between reconstruction and validation by introducing feedback mechanisms that iteratively refine the inferred topology based on consistency checks and engineering constraints. Such feedback-driven refinement has the potential to further reduce structural discrepancies and enable fully automated, trustworthy P&ID digitisation pipelines based on multimodal language models.

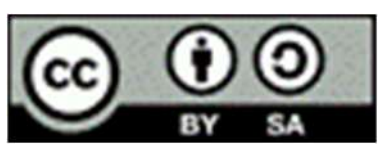